\documentclass[11pt]{article}
\usepackage{times}
\usepackage{amsmath,amssymb,amsfonts,amsthm}
\usepackage{graphicx}
\usepackage[colorlinks=true,urlcolor=blue]{hyperref}
\def\be{\begin{equation}}
\def\ee{\end{equation}}
\def\bea{\begin{eqnarray}}
\def\eea{\end{eqnarray}}
\def\lb{\label}
\def\ct{\cite}
\def\bi{\bibitem}
\begin{document}
\title{\sc How is non-knowledge represented in economic theory?}
\author{
{\sc Ekaterina Svetlova}\thanks{E--mail:
{\tt esvetlova@karlshochschule.de}}
\ and \ 
{\sc Henk van Elst}\thanks{E--mail:
{\tt hvanelst@karlshochschule.de}} \\
{\small\em ${}^{1}$Fakult\"{a}t I: Betriebswirtschaft und
Management,
Karlshochschule International University} \\
{\small\em Karlstra\ss e 36--38, 76133 Karlsruhe, Germany}}

\date{\normalsize{September 10, 2012}}
\maketitle
\begin{abstract}
In this article, we address the question of how non-knowledge 
about future events that influence economic 
agents' decisions in choice settings has been formally 
represented in economic theory 
up to date. To position our discussion within the ongoing debate 
on uncertainty, we provide a brief review of historical 
developments in economic theory and decision theory on 
the description of economic agents' choice behaviour under 
conditions of uncertainty, understood as either (i)~ambiguity, or 
(ii)~unawareness. Accordingly, we identify and discuss two 
approaches to the formalisation of non-knowledge: one based on 
decision-making in the context of a state space representing the 
exogenous world, as in Savage's axiomatisation and some 
successor concepts (ambiguity as situations with unknown 
probabilities), and one based on decision-making over a set of 
menus of potential future opportunities, providing the possibility 
of derivation of agents' subjective state spaces (unawareness as 
situation with imperfect subjective knowledge of all future events 
possible). We also discuss impeding challenges of the 
formalisation of non-knowledge.

\end{abstract}
\section{Introduction}
\lb{sec1}
The recent economic crisis once again drew attention to the 
insufficient ability of modern economic theory to properly account 
for uncertainty and imperfect knowledge: neglect of these issues 
is argued to be one of the reasons for the failure of the economic 
profession in the difficult times of 2007--2009; cf.~{\em The 
Economist\/} (2007) \ct{eco2007}, Colander {\em et al\/} (2009) 
\ct{coletal2009}, Taleb (2010) \ct{tal2010}, Akerlof and Shiller 
(2009) \ct{akeshi2009}, and Svetlova and Fiedler (2011) 
\ct{svefie2011}. Next to the voices from the inside of the 
profession, there is the related criticism from neighbouring 
disciplines such as, e.g., economic sociology; cf.~Beckert (1996) 
\ct{bec1996}, and Esposito (2007, 2010) \ct{esp2007, esp2010}. The 
impression arises that economists are utterly ignorant: they 
supposedly do not pay (enough) attention to the issues which the 
rest of the world consider to be most crucial for economic life. 
We asked ourselves if this ignorance is indeed a part of 
scientific practice in economics. Is it correct that nobody has 
properly tackled the issue of true uncertainty and imperfect 
knowledge since Knight (1921) \ct{kni1921} and Keynes (1921) 
\ct{key1921} during the post-WW~I twentieth century?

In this article, we aim to arrive at a more differentiated 
judgement. Based on a 
review of the literature, we classify the developments in 
economics and decision theory that refer to uncertainty and 
imperfect knowledge. We identify three major directions that deal 
with these issues in economics, specifically {\em risk\/}, {\em 
uncertainty as ambiguity\/}, and {\em uncertainty as 
unawareness\/}. However, it should be stressed that our goal is 
not a detailed classification of 
approaches {\em per se\/}, but answering the question of how 
{\em non-knowledge\/} has been represented formally in 
economic theory to date. This task requires, however, some 
detailed detection work, because {\em non-knowledge\/} has not 
been an explicit issue in economics yet.

Surely, there is knowledge economy, cf.~Rooney {\em et al\/} 
(2008) \ct{roo2008}, where knowledge is treated as a resource or a 
desirable asset. Also, knowledge is an important topic in 
information economics, as pioneered by Stigler (1961) 
\ct{sti1961}, Akerlof (1970) \ct{ake1970}, Spence (1973) 
\ct{spe1973}, and Stiglitz (1975, 2002) \ct{sti1975, sti2002}, 
where it is considered to be one of the tools to maximise profit. 
Generally, in economics, knowledge is considered as a good that is 
commonly available in principle (and should be used); the opposite 
--- non-knowledge --- 
is treated implicitly as a lack of information. In philosophy and 
the social sciences, the situation is not very different, though 
there are interesting recent attempts to overcome ``theoretical 
preoccupations that underlie the study of knowledge accumulation,''
McGoey (2012) \ct[p~1]{mcg2012}, and to develop an agenda for the 
social and cultural study of ignorance; cf.~McGoey (2012) 
\ct{mcg2012} and Proctor (2008) \ct{pro2008}. Ignorance should be 
treated ``as more than `not yet known' or the steadily retreating 
frontier,'' Proctor (2008) \ct[p~3]{pro2008}, and should be 
separately accounted for as a strategic resource and the source of 
economic profit and progress; cf.~Knight (1921) \ct{kni1921} and 
Esposito (2010) \ct{esp2010}. In economic theory, there have been 
occasional voices pleading for more attention to ``true 
uncertainty'', understood as the principle impossibility of 
foreseeing all future events that may occur in the exogenous 
world, cf.~Davidson (1991) \ct{dav1991} and Dequech (2006) 
\ct{deq2006}, and to ``unknown unknowns'', cf.~Taleb (2007) 
\ct{tal2007} and Diebold {\em et al\/} (2010) \ct{dieetal2010}. 
However, non-knowledge has not become an independent issue of any 
significant interest or importance for economists so far. Thus, to 
find out how ignorance is formalised in the approaches considered 
here, we have to uncover first which aspects of decision-making 
are treated (often indirectly) as unknown, and which mathematical 
instruments are used to represent them.

Our focus is on the principle non-knowledge of future events in 
the exogenous world, which is the primary source of uncertainty. 
After providing, in Section \ref{sec2}, a brief historical 
overview to position the approaches considered within the ongoing 
debate on uncertainty, we are concerned with the formal 
mathematical representation of {\em ambiguity\/} in Section 
\ref{sec3}, and of {\em unawareness\/} in Section \ref{sec4}. 
Accordingly, we identify and review two approaches to the 
formalisation of non-knowledge in the literature: one based on 
economic agents' decision-making in the context of a state space 
representing the exogenous world, as in Savage's (1954) 
\ct{sav1954} axiomatisation and some successor concepts (ambiguity 
as situations with unknown probabilities), and one based on 
decision-making over a set of menus of potential future 
opportunities, providing the possibility of derivation of agents' 
subjective state spaces (unawareness as situation with imperfect 
subjective knowledge of all future events). Due to the large 
number of papers written on this topic, we have to be selective 
and, hence, cannot provide an exhaustive overview. We particularly 
draw attention to the last-mentioned line of research, namely 
uncertainty as unawareness, as it represents an exciting attempt 
to formalise ``unknown unknowns'' by radically departing from the 
mainstream paradigm of Savage's axiomatisation. Finally, in 
Section \ref{sec5}, we discuss the impending challenges and tasks 
of formalisation of non-knowledge in economics. We believe that 
without a detailed understanding of how non-knowledge has been 
represented in economics so far, no serious research agenda for 
studying ignorance as an independent part of economic theory can 
be developed. We hope that this article provides one of the first 
useful steps towards such an agenda.

\section{Historical developments}
\lb{sec2}
Though there has not been an explicit discussion on non-knowledge 
in economic theory, this issue permanently turns up in relation to 
the topic of uncertainty. We identified three branches in the 
literature on decision-making of economic agents under conditions 
of uncertainty --- {\em risk, ambiguity\/} and {\em 
unawareness\/} --- and, in what follows, present those three 
directions and discuss the issue of knowledge versus ignorance in 
relation to each of them:
\begin{itemize}

\item[(i)] {\em risk\/}: in formal representations, possible 
states and events regarding the exogenous world and their 
respective probabilities are known to all economic agents; they 
agree on the probability measure to be employed in calculations of 
individual utility,

\item[(ii)] {\em uncertainty I -- ambiguity\/}: in formal 
representations, possible states and events are known but their 
respective probabilities are not known to the agents; each of them 
employs their own subjective (prior) probability measure in 
calculations of individual utility,

\item[(iii)] {\em uncertainty II -- unawareness\/}: in formal 
representations, possible states and events are known only 
incompletely to the agents; there is ignorance among them as 
regards relevant probability measures for calculations of 
individual utility.

\end{itemize}
This classification goes back to the work on uncertainty by Knight 
(1921) \ct{kni1921}, Keynes (1921, 1937) \ct{key1921, key1937}, 
Shackle (1949, 1955) \ct{sha1949, sha1955}, and Hayek (1945) 
\ct{hay1945}, who tightly connected 
the discussion of uncertainty with two kinds of knowledge, or 
rather ignorance: specifically, with imperfect knowledge of future 
events (uncertainty II), and with knowledge or non-knowledge of 
probability measures relating to future events (uncertainty I). 
Though the detailed depiction of the historical development of 
those concepts would go far beyond the scope of this paper, we 
consider it important to highlight the main ideas in this 
development in order to provide a topical frame for our discussion 
on the conceptualisation of non-knowledge in contemporary economic 
theory.

Generally, the authors mentioned differentiate between 
{\em epistemological\/} and {\em ontological uncertainty\/}. 
{\em Epistemological uncertainty\/} is related to situations where 
economic agents lack the knowledge necessary to construct adequate 
probability measures. According to Knight (1921) \ct{kni1921}, 
e.g., theoretical, i.e., {\em a priori\/} probabilities on the one 
hand, and statistical probabilities on the other,
are based on a valid fundament of knowledge: the law of large 
numbers, or statistical grouping. The {\em a priori\/} probability 
can be predicted using counting principles and a completely 
homogeneous classification of instances (e.g., by rolling dice), 
the statistical probability describes the frequency of an outcome 
based on a classification of empirical events or instances, given 
repeated trials. Knowledge is understood 
in both cases as (empirical) information that allows for the
classification of possible outcomes. These two kinds of 
probability ({\em a priori\/} and statistical) can be measured, 
and in this sense are known and unanimously agreed upon by all 
agents involved in decision-making processes (the situation of 
{\em risk\/}). Hence, such probability measures can be reasonably 
referred to as objective.

However, Knight suggests that these two categories do not exhaust 
all possibilities for defining a probability measure; he 
adds ``estimates'', or subjective probabilities. Quoting Knight 
(1921) \ct[p~225]{kni1921}: ``The distinction 
here is that there is no valid basis of any kind for classifying 
instances. This form of probability is involved in the greatest 
logical difficulties of all \ldots.'' Knight refers to this last 
situation as a situation of {\em uncertainty\/} (ibid 
\ct[p~233]{kni1921}); uncertainty can be defined as absence of 
probable knowledge. In the situation of risk, probabilities 
represent the measurable degree of non-knowledge; in the 
uncertainty situation, this degree is immeasurable, and in this 
sense probabilities are not known. Keynes (1921) \ct{key1921} also 
suggested a concept of immeasurable probabilities as logical 
relationships, and argued in his 1937 paper --- in unison with 
Knight --- that economic agents lack a valid basis to devise 
probability measures. In his definition uncertainty exists, e.g., 
in the case of predicting the price of copper or the interest rate 
20 years hence (Keynes (1937) \ct[p~113]{key1937}): ``About these 
matters there is no scientific basis on which to form any 
calculable probability whatever. We simply do not know.'' 
Probabilities are used by economic agents as a convention that 
enables them to act (ibid \ct[p~114]{key1937}); at the same time, 
though probabilities are widely applied, they represent the 
agents' ignorance rather than their (scientific) knowledge.

Interestingly, in the later literature this issue was taken up by 
Ellsberg (1961) \ct{ell1961}, who, in his experiments, 
distinguished between situations with known probability measures 
over some event space (when the color and number of the balls in 
an urn are known to agents; thus, they can form probabilities), 
and situations with unknown probability measures (agents know only 
the colors of balls but not the exact number of balls of each 
color; thus, they deal with the ignorance of probability). 
Ellsberg demonstrated empirically that people tend to prefer 
situations with known probability measures over situations with 
unknown probability measures; he explicitly referred to situations 
with unknown probability measures as {\em ambiguous\/} and named 
the phenomenon of avoiding such situations ``ambiguity aversion'' 
(corresponding to the term ``uncertainty aversion'' coined by 
Knight (1921) \ct{kni1921}).   

It must be noted that the discussion about measurability of 
probabilities in economic life, as well as about their objective 
vs subjective character, was severely influenced and pulled in one 
particular, for a long time uncontested, direction by the line of 
argumentation due to Ramsey (1931) \ct{ram1931}, de Finetti 
(1937) \ct{fin1937}, and Savage (1954) \ct{sav1954}. Ramsey 
and de Finetti reacted to Knight's and Keynes' concepts of 
uncertainty as situations with immeasurable probabilities with the 
axiomatisation of subjective probabilities: they demonstrated that 
subjective probabilities can always be derived from the observed 
betting behaviour of economic agents, rendering the whole 
discussion about measurability and objectivity of probabilities 
seemingly obsolete. Adopting these results, Savage generalised the 
theory of decision under risk, i.e., the expected utility theory 
as conceived of originally by Bernoulli (1738) \ct{ber1738} and 
von Neumann and Morgenstern (1944) \ct{neumor1944}. While the 
expected utility concept as an element of risk theory was based on 
objective probability measures, Savage combined expected utility 
theory and the subjective probability approach of Ramsey and de 
Finetti to deliver a new variant of an
axiomatisation of decision under conditions of uncertainty --- 
subjective expected utility theory. This concept was perfectly 
compatible with the Bayes--Laplace approach to probability theory 
and statistics where subjective {\em prior\/} probabilities can 
always be assumed to exist and adjusted in the process of 
learning. The crucial feature of Savage's probabilistic 
sophistication is the principle neglect of the Knightian 
distinction between risk and uncertainty, as Savage's concept 
presupposes that even if an objective probability measure for 
future events is not known, it can always be assumed that 
economic agents behave {\em as if\/} they apply an individual 
subjective (prior) probability measure to estimating the 
likelihood of future events; and these probability measures can in 
principle be derived {\em a posteriori\/} from an axiomatic model 
on the basis of empirical data on agents' choice behaviour. By 
this theoretical move, the immeasurability (and thus the 
knowability) issue is eliminated. The question of the validity of 
the subjective degrees of beliefs foundation, or of the origin of 
subjective probabilities, is beyond Savage's model, as these are 
built into the {\em as-if\/}-construction from the outset.

However, the Knightian distinction continued to bother economists 
and --- especially after Ellsberg's (1961) \ct{ell1961} paper 
--- a new branch of research appeared in the literature that 
endeavoured to re-introduce uncertainty, understood as absence of 
perfect knowledge of relevant probability measures, into economic 
theory. The most prominent attempt was delivered by Gilboa and 
Schmeidler (1989) \ct{gilsch1989}.
In the next section, we will introduce the basic elements of their 
axiomatisation of decision under uncertainty in terms of 
non-unique probability measures,
and contemplate how non-knowledge is represented in this concept.
At the same time, the attentive reading of Knight, Keynes and 
Shackle suggests that the issue of uncertainty is not restricted 
to the question whether probabilities can be meaningfully defined 
or measured. There is a more fundamental issue of {\em ontological 
uncertainty\/} which is concerned with the principle unknowability 
of what is going on in an economic system; it goes beyond the 
scope of epistemic uncertainty. 

Note that in the framework of epistemic uncertainty, knowledge 
that is relevant for the derivation of a meaningful probability 
measure is generally treated as information; compare the 
respective definition by Epstein and Wang (1994) 
\ct[p~283]{epswan1994}, who define risk as a situation ``where 
probabilities are available to guide choice, and uncertainty, 
where information is too imprecise to be summarized adequately by 
probabilities.'' It is interesting that also beyond the borders of 
economic theory --- in the IPCC (2007) \ct{ipcc2007} report
--- the Knightian distinction between risk and uncertainty is 
understood as an epistemic one: ``The fundamental distinction 
between `risk' and `uncertainty' is as introduced by economist 
Frank Knight (1921), that risk refers to cases for which the 
probability of outcomes can be ascertained through 
well-established theories with reliable complete data, while 
uncertainty refers to situations in which the appropriate data 
might be fragmentary or unavailable.'' (\ldots) The clear relation 
``information (empirical data) -- probabilities'' is presupposed. 
The lack of knowledge, in this case, can be theoretically removed 
by becoming more skillful in calculating, or by collecting more 
information. 

However, it should be stressed that Knight (as well as Keynes and 
Shackle) did not conceive of ignorance as lack of information but 
rather as ontological indeterminacy, the ``inherent unknowability 
in the factors'', see Knight (1921) \ct[p~219]{kni1921}. Shackle 
(1955) \ct{sha1955} relates the genuinely imperfect knowledge 
about future events to the 
absence of an exhaustive list of possible consequences of choices. 
Traditional probability theory assumes that the list of 
consequences over which probability is distributed is an 
exhaustive list of possible outcomes, or, in Shackle's terms,
hypotheses. However, so Shackle, if there is a residual 
hypothesis, that is, the list of possible consequences is 
incomplete, the probability model runs into trouble. By adding a 
hypothesis to the list of possible hypotheses, each corresponding 
probability of the previously known hypotheses has to be revised 
downwards; see Shackle (1955) \ct[p~27]{sha1955}. If five possible 
hypotheses are considered and a sixth hypothesis is added, and 
additivity of probabilities is assumed, the probability of each of 
the initial five hypotheses is subsequently lower.
This objection applies to both approaches, namely the frequentist 
approach to probability theory on the one hand, and the 
Bayes--Laplace approach which deals with belief-type subjective
(prior) probability measures on the other, because neither can 
incorporate a residual hypothesis, or the principle non-knowledge 
of future states. Thus, referring to the genuinely imperfect 
knowledge about future events, Shackle (but also Knight and 
Keynes) expressed doubts whether probability theory in general is 
sufficient to account for decision under uncertainty, and whether 
it should be the central issue after all.

By far more important than the issue of devising suitable 
probability measures seems to be the non-knowledge of possible 
future states of the exogenous world and of related outcomes. Only 
if we manage to account properly for this imperfect knowledge, can 
we conceptualise properly human decision-making, or, in the words 
of Shackle (1959) \ct[p~291]{sha1959}, a non-empty decision. 
Crocco (2002) \ct{cro2002} explains: ``An empty decision is the 
mere account of a formal solution to a formal problem. It is that 
situation where a person has a complete and certain knowledge 
about all possible choices and all possible outcomes of each 
choice. It is a mechanical and inevitable action," or, in the 
words of Heinz von F\"orster (1993) \ct[p~153]{foe1993}, every 
decidable (or perfectly known) problem is already decided; true 
decisions always presuppose genuine undecidability.
In this sense, Savage's concept is rather concerned with empty 
decisions, because it presupposes situations with full knowledge 
of possible events, acts and outcomes, rendering agents' choices 
just a mechanical application of the personal utility-maximisation 
rule.

In economics, genuine undecidability should enter theory. 
Most economic decisions are truly undecidable because they take 
place under conditions of imperfect knowledge of the situation to 
be faced, which is in the sense of the American pragmatist 
philosopher John Dewey (1915) \ct[p~506]{dew1915} a genuinely 
``incomplete situation'': ``something is `there', but what is 
there does not constitute the entire objective situation.'' This 
``means that the decision-maker does not have complete knowledge 
of the following: (a)~the genesis of the present situation, 
(b)~the present situation itself, or (c)~the future outcomes that 
remain contingent on the decisions that are made in the present 
situation;'' see Nash (2003) \ct[p~259]{nas2003}. According to 
Dewey (1915) \ct{dew1915}, the situation is underdetermined, 
unfinished, or not wholly given.

This principle non-knowledge can be explained, so Shackle (1949, 
1955) \ct{sha1949, sha1955}, by the character of economic 
decisions, which he considers to be non-devisible, non-seriable, 
and crucial experiments. {\em Non-devisible experiments\/} imply 
only a single trial; {\em non-seriable experiments\/} are not 
statistically important even in the aggregate; an example of a 
seriable experiment is fire insurance: although no reasonable
probability can be assigned to an individual house to burn down, 
if there are sufficiently many events, a (statistical) 
probability will emerge. Most importantly, economic decisions are 
{\em crucial experiments\/}: they inevitably alter the conditions 
under which they were performed (this definition applies to all 
strategic situations, e.g., chess play, but also financial 
markets). Within 
the genuinely social context of economic life, economic events are 
rather {\em endogenous\/} to the decision processes of agents and 
are dependent on the actions and thinking of other market 
participants. There are path dependencies and reflexivity; cf. 
Soros (1998) \ct{sor1998}. In general, a meaningful approach to 
decision-making should take into account that the future is 
principally unknowable, due to ontological features of the 
exogenous world such as openness, organic unity, and 
underdeterminacy. These are features which are typically 
attributed to complex systems; cf.~Keynes {\em et 
al\/} (1926) \ct[p~150]{keyetal1926}: ``We are faced at every turn 
with the problems of Organic Unity, of Discreteness, of 
Discontinuity --- the whole is not equal to the sum of the parts, 
comparisons of quantity fail us, small changes produce large 
effects, the assumptions of a uniform and homogeneous continuum 
are not satisfied.''
In such a system, not all constituent 
variables and structural relationships connecting them are known 
or knowable. Thus, in an open and organic system, some information 
is not available at the time of decision-making, and cannot be 
searched, obtained or processed in principle. Surprises, or 
unforeseen events, are normal, not exceptional. The list of 
possible events or states is not predetermined and very little, or 
nothing at all, can be known about the adequate probability 
measure for this radically incomplete set of future events.   

These considerations require a more sophisticated distinction of 
decision-making configurations, namely a distinction that goes 
beyond the usual {\em risk\/} vs {\em uncertainty as ambiguity\/} 
debate. As Dequech (2006) \ct[p~112]{deq2006} puts it:
``Even though the decision-maker under ambiguity does not know 
with full reliability the probability that each event (or state of 
the world) will obtain, he/she usually knows all the possible 
event \ldots. Fundamental uncertainty, in contrast, is 
characterized by the possibility of creativity and 
non-predetermined structural change. The list of possible events 
is not predetermined or knowable ex ante, as the future is yet to 
be created.'' What Dequech calls ``fundamental uncertainty'' (or 
``true uncertainty'' in terms of some post-Keynesians (e.g., 
Davidson (1991) \ct{dav1991}) enters the recent debate in the 
economic literature under the label of ``unawareness''.

The {\em unawareness\/} concept, as introduced by Kreps 1979 
\ct{kre1979}, Dekel {\em et al\/} (1998, 2001) 
\ct{deketal1998,deketal2001}, and Epstein {\em et al\/} (2007) 
\ct{epsetal2007}, presupposes a coarse (imperfect) subjective 
knowledge of all possible future events. This concept criticises 
Savage's (1954) \ct{sav1954} axiomatisation and suggests a radical 
departure from it. Savage's axiomatisation is characterised by the 
in principle observability and knowability of all possible future 
events. These events belong to the primitives of the model 
and are assumed to be exogenous and known to 
all economic agents. In Savage's model, the 
(compact) state space representing the exogenous world the
agents are continually interacting with is ``a space of 
mutually exclusive and exhaustive states of nature, representing 
all possible alternative unfoldings of the world''; see Machina 
(2003) \ct[p~26]{mac2003}. The exhaustiveness criterion is very 
restrictive and basically precludes non-knowledge of future states 
on the part of the agents. Machina (2003) \ct[p~31]{mac2003} 
continues: ``When the decision maker has reason to `expect the 
unexpected' [or the residual hypothesis in terms of Shackle --- 
the authors], the exhaustivity requirement cannot necessarily be 
achieved, and the best one can do is specify a final, catch-all 
state, with a label like `none of the above', and a very 
ill-defined consequence.'' Obviously, true uncertainty as 
imperfect knowledge of possible future states of the exogenous 
world is not an element of Savage's model.
The pioneers of the {\em unawareness\/} concept depart from 
Savage's axiomatisation by replacing the state space in the list 
of primitives by a set of menus over actions which are the objects 
of choice. This theoretical move allows for dealing with 
unforeseen contingencies, i.e., an inability of economic agents to 
list all possible future states of the exogenous world.

We now turn to give a more formal presentation of the two main 
concepts of uncertainty we discussed so far: uncertainty as 
ambiguity and uncertainty as unawareness.

\section{Uncertainty as ambiguity: non-knowledge of probability 
measures}
\lb{sec3}
All decision-theoretical approaches to modelling an economic 
agent's state of knowledge regarding future developments of the 
exogenous world,
the ensuing prospects for an individual's opportunities, and the 
agent's consequential choice behaviour under conditions of 
uncertainty employ an axiomatic description of the characteristic 
properties of observable choice behaviour and derive a 
quantitative representation of an agent's preferences in 
decision-making. Uncertainty in this context is generally 
interpreted as ambiguity perceived by an agent with respect to 
unknown probabilities by which future states of the exogenous 
world will be realised. In these approaches the standard 
assumption of neoclassical economics of an agent whose 
choices are fully rational is being maintained. The main issue 
of modelling here is to put forward a set of primitives which can 
be observed in principle in real-life settings, as well as a 
minimal set of axioms describing exhaustively the interconnections 
between these primitives, to provide the conceptual basis for (in 
general highly technically demanding) mathematical proofs of 
representation theorems. 
Most approaches in the literature propose an expected utility (EU)
representation of an agent's preferences in terms of a
real-valued personal utility function which is an unobservable 
theoretical construct, thus following the quantitative 
game-theoretical tradition of von Neumann and Morgenstern (1944) 
\ct{neumor1944}. A related issue is the question to what extent an 
agent's choice behaviour can be reasonably viewed as influenced by 
a set of personal subjective probabilities regarding the (unknown) 
future states of the exogenous world. We begin by briefly 
reviewing the central aspects of the axiomatic approach taken by 
Savage (1954) \ct{sav1954} to describe one-shot choice situations 
--- the subjective expected utility (SEU) framework, which 
attained the prominent status of a standard model in decision 
theory.

The primitives in Savage (1954) \ct{sav1954} are
\begin{itemize}
\item[(i)] an exhaustive set of mutually exclusive future states 
$\omega$ of the exogenous world which an agent cannot actively
take an influence on; these constitute a state space 
$\boldsymbol{\Omega}$ which is assumed to be continuous,
compact, and can be partitioned into a finite number of pairwise 
disjoint events; possible events $A, B, \ldots$ are considered 
subsets of $\boldsymbol{\Omega}$, with $2^{\boldsymbol{\Omega}}$ 
the set of all such subsets of $\boldsymbol{\Omega}$,

\item[(ii)] a finite or infinite set of outcomes $x$ contingent on 
future states $\omega$, forming an outcome space 
$\boldsymbol{X}$, and

\item[(iii)] a weak binary preference order $\succeq$ (``prefers 
at least as much as'') defined over the agent's objects of choice 
--- a set of potential individual acts $f$ an agent may 
consciously take in reaction to realised future states~$\omega$ of 
the exogenous world, yielding predetermined outcomes $x$ ---,
describing their personal ranking of available options; these acts 
form a space $\boldsymbol{F}$.
\end{itemize}
In more detail, an act is defined as a (not necessarily 
real-valued, continuous) mapping $f: \boldsymbol{\Omega} 
\rightarrow \boldsymbol{X}$ from the set of future states 
$\boldsymbol{\Omega}$ to the set of possible outcomes 
$\boldsymbol{X}$, so the set of acts available to an agent at 
a given instant in time, in view of known future states $\omega$ 
but of unknown probabilities, is 
$\boldsymbol{F} = \boldsymbol{X}^{\boldsymbol{\Omega}}$.
There is no additional structure needed in this model regarding 
measures or topology on either space $\boldsymbol{\Omega}$ or 
$\boldsymbol{X}$, except for continuity and compactness of 
$\boldsymbol{\Omega}$. An observable weak binary preference order 
over the set of acts is given by $\succeq \subset \boldsymbol{F} 
\times \boldsymbol{F}$, intended to reflect an agent's subjective 
beliefs regarding future states $\omega$, and the usefulness of 
acts the agent may take in response to ensuing states.

Savage introduces a minimal set 
of seven axioms (P1 to P7) to characterise the theoretical nature 
of this preference order over acts (and, by implication,
related outcomes), which are commonly referred to in the 
literature as weak order resp.~completeness, sure-thing principle, 
state-independence, comparative probability, non-triviality, 
Archimedean, and finitely additive probability measures; cf. 
Nehring (1999) \ct[p~105]{ner1999} and Gilboa (2009) 
\ct[p~97ff]{gil2009}. These 
axioms constitute the foundation of a representation theorem 
proved by Savage which states that an agent's (one-shot) choice 
behaviour 
under conditions of uncertainty may be viewed as if it was guided 
by (i)~a real-valued personal utility function $U: \boldsymbol{X} 
\rightarrow \mathbb{R}$ that assigns subjective value to specific 
outcomes $x \in \boldsymbol{X}$, and (ii)~a single finitely 
additive subjective probability measure $\mu: 
2^{\boldsymbol{\Omega}} \rightarrow [0,1]$ on the space of all 
possible future events $2^{\boldsymbol{\Omega}}$. In particular, 
an agent's choice behaviour may be modelled as if for the acts $f$ 
available to them they strive to maximise a real-valued EU 
preference function $V: \boldsymbol{F} \rightarrow 
\mathbb{R}$, defined by
\be
\lb{eq:savagerepr}
V(f) := \int_{\boldsymbol{\Omega}}U(f(\omega))\,\mu({\rm d}\omega) 
\ .
\ee
Hence, in this setting an act $f \in \boldsymbol{F}$ is weakly 
preferred by an agent to an act $g \in \boldsymbol{F}$, iff
$V(f) \geq V(g)$.

The elements of Savage's SEU model may be schematically summarised 
in terms of a decision matrix of the following structure (here for 
a partition of the continuous and compact $\boldsymbol{\Omega}$ 
into a finite number $n$ of pairwise disjoint events):
\be
\begin{array}{c|cccc|c}
\text{probability measure}\ \mu & P(\omega_{1}) & P(\omega_{2}) & 
\ldots & P(\omega_{n}) & \\
\hline
\text{acts}\ \boldsymbol{F}\ \backslash\ \text{states}
\ \boldsymbol{\Omega} & \omega_{1} & 
\omega_{2} & \ldots & \omega_{n} & \\
\hline
f_{1} & x_{11} & x_{12} & \ldots & x_{1n} & \\
f_{2} & x_{21} & x_{22} & \ldots & x_{2n} & \text{outcomes}
\ \boldsymbol{X} \\
\vdots & \vdots & \vdots & \ddots & \vdots & 
\end{array} \ ,
\ee
where $0 \leq P(\omega_{i}) \leq 1$ and $\sum_{i}P(\omega_{i})=1$ 
(and generally: $\mu \geq 0$ and 
$\int_{\boldsymbol{\Omega}}\mu({\rm d}\omega) = 1$). Note that, 
formally, Savage's framework reduces an 
agent's situation of decision under uncertainty, in the Knightian 
sense of not knowing the probability measure associated with 
$(\boldsymbol{\Omega}, 2^{\boldsymbol{\Omega}})$ {\em a priori\/}, 
to a manageable situation of decision under risk by introducing a 
{\em single\/} subjective Bayesian prior probability measure as a 
substitute. This is to say, every single economic agent possesses 
for themselves a unique probability measure which they employ in 
their individual calculations of utility; a probability measure is 
thus {\em known\/} to every individual from the outset, but there 
is no reason whatsoever that these measures should coincide 
between agents.

Savage's main claim is that his framework can be used to 
explicitly derive for an arbitrary economic agent who makes 
rational choices in parallel (i)~a unique subjective probability 
measure $\mu$ over $(\boldsymbol{\Omega}, 
2^{\boldsymbol{\Omega}})$, and (ii)~a personal utility function 
$U$ over $\boldsymbol{F}$ (unique up to positive linear 
transformations), from observation of their choice behaviour 
in practice. For the sequel it is worth mentioning that Savage's 
numerical SEU representation~(\ref{eq:savagerepr}) can be 
interpreted to fall into either of the categories of ordinal or 
additive EU representations.

\medskip
Various authors have criticised Savage's SEU model for different
reasons, where in particular the claim is that one or more of his 
axioms are regularly being violated in real-life situations of 
(one-shot) choice. Bewley (1986,2002) \ct{bew1986, bew2002}, for 
example, points the finger 
to the completeness axiom P1 in that he considers it unrealistic 
to assume that all agents have a clear-cut ranking of all the acts 
available to them, when it need not necessarily be clear from the 
outset which acts comprise the complete set $\boldsymbol{F}$. In 
his work he therefore proposes an axiomatic alternative to 
Savage's SEU model which discards the completeness axiom in favour 
of an inertia assumption regarding the status quo of an agent's 
personal situation.

More prominent still is Ellsberg's (1961)\ct{ell1961} empirical 
observation that in situations of choice under uncertainty 
rational agents need not necessarily act as
subjective expected utility maximisers: given the choice between a 
game of chance with known probabilities of the possible outcomes 
and the identical game of chance where the probabilities are 
unknown, the majority of persons tested exhibited the phenomenon 
of uncertainty aversion by opting for the former game. Ellsberg 
showed that this kind of behaviour correspond to a violation of 
Savage's sure-thing principle axiom P2.

\medskip
A possible resolution of 
this conflict was suggested in the multiple priors maxmin expected 
utility (MMEU) model due to Gilboa and Schmeidler (1989) 
\ct{gilsch1989}, which takes uncertainty aversion explicitly into 
account by stating that under conditions of uncertainty an agent 
need not have to have a unique subjective prior probability 
measure $\mu$, but rather an {\em entire set\/} $\Pi$ worth of such
measures $\pi$ from which they select in making decisions 
according to the maxmin principle. In this sense, Gilboa and 
Schmeidler take an explicit attempt at formalising Knightian 
uncertainty in problems of decision-making, interpreted as 
situations with in principle unknowable probability measures over 
$(\boldsymbol{\Omega},2^{\boldsymbol{\Omega}})$. The degree of an 
agent's ignorance is encoded in the generically unconstrained 
cardinality of the set of Bayesian priors~$\Pi$: no criteria are 
formulated according to which an agent assesses the relevance of 
any particular probability measure that is conceivable for a given 
situation of decision-making. Non-knowledge regarding the 
likelihood of future events here is linked to the number of 
elements included in the individual set $\Pi$ that is employed in  
an agent's individual calculation of utility and so is represented 
in a more comprehensible fashion than in Savage's framework.

Nevertheless, the primitives of the MMEU model are unchanged with 
respect to Savage's SEU model. Based on a minimal set of six 
axioms (A1 to A6) referred to resp.~as weak order, 
certainty-independence, continuity, monotonicity, 
uncertainty aversion and non-degeneracy, the representation 
theorem Gilboa and Schmeidler (1989) \ct{gilsch1989} prove employs 
a real-valued preference function $V: \boldsymbol{F} \rightarrow 
\mathbb{R}$ defined by the minimum expected utility relation
\be
\lb{eq:gilschrepr}
V(f) := \min_{\pi\in \Pi}
\int_{\boldsymbol{\Omega}}(E_{f(\omega)}U)\,{\rm d}\pi \ ,
\ee
with $\Pi \subset \Delta(\boldsymbol{\Omega})$ a non-empty, 
closed and convex set of finitely additive probability measures 
over $(\boldsymbol{\Omega},2^{\boldsymbol{\Omega}})$, and $U: 
\boldsymbol{X} \rightarrow \mathbb{R}$ a non-constant real-valued 
personal utility function. Again, an act $f \in \boldsymbol{F}$ is 
then weakly preferred by an agent to an act $g \in 
\boldsymbol{F}$, iff $V(f) \geq V(g)$.

Since its inception, Gilboa and Schmeidler's MMEU model has 
enjoyed a number of applications in the econometrical literature; 
e.g. in Epstein and Wang (1994) \ct{epswan1994} on intertemporal 
asset pricing; Hansen {\em et al\/} (1999) \ct{hanetal1999} on 
savings behaviour; Hansen and Sargent (2001, 2003) 
\ct{hansar2001,hansar2003} on macroeconomic situations; 
Nishimura and Ozaki (2004) \ct{nisoza2004} on a job search model; 
and Epstein and Schneider (2010) \ct{epssch2010} on implications 
for portfolio choice and asset pricing. Rigotti and Shannon (2005) 
\ct{rigsha2005}, who propose an approach to formalising 
uncertainty in financial markets on the basis of Bewley's 
(1986,2002) \ct{bew1986,bew2002} idea of discarding Savage's 
completeness axiom P1, contrast their findings on the impact of 
uncertainty on equilibrium configurations in decision-making 
processes with corresponding consequences arising from an MMEU 
perspective.

\medskip
The strongest criticism to date of Savage-type state space models 
of decision-making under conditions of uncertainty was voiced at 
the end of the 1990ies by Dekel {\em 
et al\/} (1998) \ct{deketal1998}. They showed that given one 
considers it unrealistic for an economic agent to be aware of all 
possible future states $\omega$ of the exogenous world, a standard 
state space model is incapable of consistently incorporating the 
dimension of an agent's unawareness of future contingencies. 
The basis of the formal treatment of the issue at hand are 
information structures referred to as possibility  
correspondences.
A possibility correspondence amounts to a function $P: 
\boldsymbol{\Omega} \rightarrow 
2^{\boldsymbol{\Omega}}$ that maps elements~$\omega$ in some state 
space $\boldsymbol{\Omega}$ to subsets thereof, so that 
$P(\omega)$ is interpreted as the set of states an agent considers 
possible when the realised state is~$\omega$. In this picture, an 
agent ``knows'' an event $E \in 
2^{\boldsymbol{\Omega}}$ at a state~$\omega$ provided $P(\omega) 
\subseteq E$. Hence, given a possibility correspondence $P$, a 
knowledge operator $K: 2^{\boldsymbol{\Omega}} \rightarrow 
2^{\boldsymbol{\Omega}}$ is determined by
\be
K(E) := \{\omega \in \boldsymbol{\Omega}|P(\omega) \subseteq E\}
\quad\text{for all}\quad
E \in 2^{\boldsymbol{\Omega}} \ ;
\ee
$K(E)$ represents the set of states in $\boldsymbol{\Omega}$ for 
which an agent knows that event $E$ must have occurred.
According to Dekel {\em et al\/}, it is commonplace to assume that 
such a knowledge operator features the properties of 
(i)~necessitation, meaning 
$K(\boldsymbol{\Omega})=\boldsymbol{\Omega}$, and 
(ii)~monotonicity, meaning $E \subseteq F \Rightarrow K(E) 
\subseteq K(F)$. In addition, an unawareness operator may be 
defined as a mapping $U: 2^{\boldsymbol{\Omega}} \rightarrow 
2^{\boldsymbol{\Omega}}$, so that $U(E)$ is to be regarded as the 
set of states in $\boldsymbol{\Omega}$ where an agent is unaware 
of the possibility that event $E$ may occur. With these structures 
in place, a standard state space model is represented by a triplet 
$(\boldsymbol{\Omega},K,U)$.

To obtain their central result, Dekel {\em et al\/} require a 
minimal set of only three axioms which characterise the nature of 
the operators $K$ and $U$: these demand that for every event $E 
\in 2^{\boldsymbol{\Omega}}$, (i)~$U(E) \subseteq \neg K(E) \cap 
\neg K(\neg K(E))$, called plausibility,\footnote{The symbol 
$\neg$ denotes complementation.} (ii)~$K(U(E)) = 
\emptyset$, called KU introspection, and (iii)~$U(E) \subseteq 
U(U(E))$, called AU introspection. Given a standard state space 
model $(\boldsymbol{\Omega},K,U)$ satisfies these three axioms, 
the theorem proven by Dekel {\em et al\/} (1998) 
\ct[p~166]{deketal1998} states that in such a setting 
(a)~``the agent is never unaware of anything,'' provided $K$ 
satisfies the necessitation property, and (b)~``if the agent is 
unaware of anything, he knows nothing,'' provided $K$ satisfies 
the monotonicity property. This result renders standard state 
space models void as regards the intention of formally capturing 
an agent's unawareness of subjective contingencies in a 
non-trivial way.

\medskip
The work by Dekel {\em et al\/} (1998) \ct{deketal1998}, in 
particular, triggered a series of papers written during the last 
decade, which aspire to include an agent's unawareness of future 
subjective contingencies in a coherent model that continues to 
employ a kind of EU representation of an agent's manifested 
preferences in situations of choice under conditions of 
uncertainty. We turn to highlight the, in our view, most important 
papers of this development next.

\section{Uncertainty as unawareness: non-knowledge of complete 
state spaces}
\lb{sec4}
Since the status of possible future states $\omega$ of the 
exogenous world as a primitive in a decision-theoretical model on 
an agent's choice behaviour under conditions of uncertainty is 
questionable due to the lack of a convincing operational 
instruction for observation of such states, a number of authors 
have dropped the state space $\boldsymbol{\Omega}$ from the set of 
primitives altogether and turned to focus instead on the 
description of an agent's preferences when they are unaware of 
some future subjective contingencies which take a direct influence 
on future outcomes such as the pay-offs of certain actions. In the 
papers to be considered in the following, the conceptual line of 
thought pursued in which originated in the work by Kreps (1979) 
\ct{kre1979}, the primitives underlying this alternative approach
comprise in general
\begin{itemize}
\item[(i)] a (typically finite) set $\boldsymbol{B}$ of 
alternative opportunities, actions, or options; a generic element 
in this set will be denoted by $b$,

\item[(ii)] a (typically finite) set $\boldsymbol{X}$ of all 
conceivable non-trivial menus compiled from elements in 
$\boldsymbol{B}$, with a generic element denoted by $x$; note that 
$\boldsymbol{X} = 2^{\boldsymbol{B}}\backslash\{\emptyset\}$,

\item[(iii)] a weak binary preference order $\succeq$ defined over 
the agent's objects of choice, presently menus in $\boldsymbol{X}$.
\end{itemize}
The setting conceived of in this approach considers a two-stage 
choice process in which an agent will initially (``now'') choose a 
particular menu $x$, from which, contingent on subsequently 
ensuing states $\omega$ of the exogenous world, they will choose a 
specific element $b$ at an unmodelled later stage 
(``then'').\footnote{As will be described in the following, in 
some of the works to be reviewed the elements of choice at stage 
``then'' can be more complex objects than simply elements $b \in 
\boldsymbol{B}$.} Hence, two kinds of (weak) binary preference 
orders need to be introduced: an ``ex ante preference'' 
(preference ``now'') over the set $\boldsymbol{X}$, $\succeq 
\subset \boldsymbol{X} \times \boldsymbol{X}$, and an ``ex post 
preference'' (preference ``then'') over $\boldsymbol{B}$ 
contingent on a realised state $\omega$, 
$\succeq^{*}_{\omega} \subset \boldsymbol{B} \times 
\boldsymbol{B}$; cf.~Dekel {\em et al\/} (2001) \ct{deketal2001}. 
Generally, authors then proceed to formulate 
minimal sets of axioms for the ex ante preference order $\succeq$, 
on the basis of which they prove  representation theorems for 
modelling an agent's choice behaviour 
under conditions of uncertainty in the sense that the agent is 
unaware of some future subjective contingencies. A particularly 
interesting feature of some of the works to be discussed in the 
sequel is the possibility to derive in principle an agent's 
subjective state space regarding future subjective contingencies
from observed choice behaviour, given some form of
EU representation of the agent's preference relation is 
employed. This aspect is key to a meaningful representation of 
non-knowledge in economic theory. It is also seen as an 
intermediate step towards derivation of an agent's subjective 
probability measure regarding choice behaviour under conditions of 
uncertainty on the basis of empirical data.

\medskip
Kreps (1979) \ct{kre1979}, in his pioneering paper, considers an 
agent with a ``desire for flexibility'' as regards 
decision-making, the choice behaviour of which, however, may {\em 
not\/} satisfy ``revealed preference''. He formalises these
properties of an agent's envisaged choice behaviour in terms of 
the following two axioms: for all $x, x^{\prime}, x^{\prime\prime} \in 
\boldsymbol{X}$,
\be
\lb{eq:flexibility}
x \supseteq x^{\prime}
\quad\Rightarrow\quad
x \succeq x^{\prime} \ ,
\ee
%
and
\be
\lb{eq:revpref}
x \sim x \cup x^{\prime}
\quad\Rightarrow\quad
x \cup x^{\prime\prime} \sim x \cup x^{\prime} \cup 
x^{\prime\prime} \ ,
\ee
with $\sim$ denoting the indifference relation on $\boldsymbol{X}$.
Note that in the literature the axiom (\ref{eq:flexibility}) is 
often referred to as the monotonicity axiom. Kreps, in his 
discussion, does {\em not\/} make explicit an agent's uncertainty 
regarding unawareness of (some) future subjective contingencies. 
Rather, it is implied by the agent's ``desire for flexibility''. 
He continues to prove that, given a ``dominance relation'' on 
$\boldsymbol{X}$ defined by
\be
x \geq x^{\prime}
\quad\text{if}\quad
x \sim x \cup x^{\prime} \ ,
\ee
and the axioms stated before, an agent's 
preferences on $\boldsymbol{X}$ can be sensibly described as if 
they were ``maximizing a `state dependent utility function of 
subsequent consumption'{}'' in terms of a formal real-valued 
preference function $V: \boldsymbol{X} \rightarrow \mathbb{R}$,
defined by
\be
V(x) := \sum_{s \in \boldsymbol{S}}\max_{b \in x} U(b,s) \ .
\ee
Here $\boldsymbol{S}$ denotes the unobservable finite subjective 
state space of an agent's personal tastes, with generic element 
$s$, and $U: \boldsymbol{B} \times \boldsymbol{S} \rightarrow 
\mathbb{R}$ is the agent's unobservable state-dependent 
real-valued utility function of alternative opportunities 
available in the finite set $\boldsymbol{B}$. Kreps points out 
that this representation is principally ordinal in character.
The bottom-line of Kreps' approach is that the set of 
state-dependent ex post utilities $\{U(\cdot,s)|s \in 
\boldsymbol{S}\}$, expressing the agent's beliefs on potential 
future pay-offs, can be interpreted as an agent's implicitly 
given coherent subjective state space which describes their 
uncertainty regarding ex post choices over the set 
$\boldsymbol{B}$, and so can be legitimately used as a model of 
unforeseen contingencies (cf.~Kreps (1992)~\ct{kre1992}).

\medskip
However, as Dekel {\em et al\/} (2001) \ct[p~892, 
p~896f]{deketal2001} emphasise, Kreps' implied subjective state 
space $\{U(\cdot,s)|s \in \boldsymbol{S}\}$ of an  agent is far 
from being determined uniquely, since 
the axioms he proposed prove not to be sufficiently restrictive 
for this purpose. It is this  feature in particular, which these 
authors set out to overcome in their own work. To accomplish this 
goal, Dekel {\em et al\/} (2001) \ct{deketal2001} extend Kreps' 
analysis in two respects. On the one-hand side, here the agent's 
objects of choice are, in the spirit of von Neumann 
and Morgenstern (1944) \ct{neumor1944}, sets of lotteries 
$\Delta(\boldsymbol{B})$ defined over finite sets of future 
alternative opportunities $\boldsymbol{B}$, on the other, the 
assumption of an agent's strict preference for flexibility is 
relaxed to also allow for a preference for commitment in instances 
when this appears valuable. The latter feature introduces the 
possibility of an agent's view ``ex ante'' to differ from their 
view ``ex post''. To continue with the primitives: Dekel 
{\em et al\/} take the set $\Delta(\boldsymbol{B})$ to correspond 
to a set of probability measures over $\boldsymbol{B}$; a generic 
lottery in $\Delta(\boldsymbol{B})$ is denoted by $\beta$. Subsets 
of $\Delta(\boldsymbol{B})$ are referred to as menus $x$, with 
$\boldsymbol{X}$ denoting the set of all non-empty subsets of 
$\Delta(\boldsymbol{B})$. $\boldsymbol{X}$ is endowed with a 
Hausdorff topology and constitutes the formal basis of an agent's 
binary ex ante preference order, $\succeq \subset 
\boldsymbol{X} \times \boldsymbol{X}$. The two-stage choice 
process of Kreps (1979) \ct{kre1979} remains qualitatively 
unchanged: the agent chooses a menu $x \in \boldsymbol{X}$ 
``now'', and a lottery $\beta \in x$ ``then''.

Dekel {\em et al\/}'s different kinds of representations of an 
agent's ex ante preference order $\succeq$ over menus $x$ of 
lotteries correspond to triplets $(\boldsymbol{\Omega},U,u)$, 
comprising the following three common elements: a non-empty 
(exogenous) state space $\boldsymbol{\Omega}$ 
serving merely as an index set to label ex post preferences
over $\Delta(\boldsymbol{B})$, a state-dependent real-valued 
personal utility function $U: \Delta(\boldsymbol{B}) \times 
\boldsymbol{\Omega} \rightarrow \mathbb{R}$, and a real-valued 
personal aggregator function $u: \mathbb{R}^{\boldsymbol{\Omega}} 
\rightarrow \mathbb{R}$. The aggregator function is a rather 
special feature in Dekel {\em et al\/}'s analysis. It is given the 
role of translating an agent's ex post utility levels of menus 
$x$ into corresponding ex ante 
values, making the strong assumption that, in the model proposed, 
an agent has a coherent view of all future utility possibilities 
of menus $x$ available to them. The ex post preference order 
$\succeq^{*}_{\omega}$ over $\Delta(\boldsymbol{B})$, given a 
state $\omega \in \boldsymbol{\Omega}$, can be viewed as being 
encoded in the utility function $U(\cdot,\omega)$. In consequence, 
Dekel {\em et al\/} define an agent's subjective state space as 
the set $\boldsymbol{P}(\boldsymbol{\Omega},U) :=  
\{U(\cdot,\omega)|\omega \in \boldsymbol{\Omega}\}$.

On the basis of an ex ante preference-characterising minimal 
set of seven axioms (A1 to A7), referred to as weak order, 
continuity, non-triviality, indifference to randomisation (IR), 
independence, weak independence and monotonicity, resp., Dekel 
{\em et al\/} (2001) \ct{deketal2001} prove existence
theorems for three kinds of EU representations, 
all of which can be cast in the form of a real-valued preference 
function $V: \boldsymbol{X} \rightarrow \mathbb{R}$ defined by
\be
V(x) := u\left(\left(\sup_{\beta \in x}
U(\beta,\omega)\right)_{\omega \in \boldsymbol{\Omega}}\right) \ ,
\ee
with $U(\cdot,\omega)$ an EU affine function in line with von 
Neumann and Morgenstern (1944) \ct{neumor1944}, i.e., 
for all $\beta \in \Delta(\boldsymbol{B})$ and $\omega \in 
\boldsymbol{\Omega}$,
\be
U(\beta,\omega) := \sum_{b \in \boldsymbol{B}}\beta(b) 
U(b,\omega) \ .
\ee
The main results following from the proofs of the EU 
representation theorems for the binary ex ante preference
order $\succeq$ are: (i)~uniqueness of an agent's subjective state 
space $\boldsymbol{P}(\boldsymbol{\Omega},U)$ related to 
their binary ex post preference order, as well as essential 
uniqueness of the associated aggregator function $u$, (ii)~the 
size of an agent's subjective state space 
$\boldsymbol{P}(\boldsymbol{\Omega},U)$ can be interpreted as a 
measure of their uncertainty about future subjective 
contingencies, while the associated aggregator $u$ indicates 
whether such contingencies trigger a preference for commitment 
or rather for flexibility, (iii)~ordinal EU representations offer 
the smallest subjective state space 
$\boldsymbol{P}(\boldsymbol{\Omega},U)$ possible for any ordinal
representation, and (iv)~existence of an additive EU 
representation when in particular the standard independence axiom
due to von Neumann and Morgenstern (1944) \ct{neumor1944} holds; 
the former is given by a real-valued preference function 
$V: \boldsymbol{X} \rightarrow \mathbb{R}$ such that (up to 
monotone transformations)
\be
V(x) = \int_{\boldsymbol{\Omega}}\sup_{\beta \in x}
U(\beta,\omega)\,\mu({\rm d}\omega) \ ,
\ee
with $\mu$ a (non-unique) finitely additive probability measure on 
$(\boldsymbol{\Omega}, 2^{\boldsymbol{\Omega}})$.
This last result, providing a representation in line with 
``standard'' approaches, raises ideas on the possibility of 
identification of an agent's probability measure over their 
subjective state space $\boldsymbol{P}(\boldsymbol{\Omega},U)$, 
analogous to one of the central outcomes of Savage's (1954) 
\ct{sav1954} SEU model. However, it is the state-dependence of an 
agent's ex post preference which renders this objective 
currently quite unrealistic. 

Dekel {\em et al\/}'s (2001) \ct[p~894]{deketal2001} approach 
contains an inherent interpretational difficulty, which these 
authors briefly address: the model represents an agent with an 
at least partially incomplete concept of future subjective 
contingencies ``now'' with an agent with complete knowledge 
of all utility possibilities of menus ``then''; does the 
model, nevertheless, deal consistently with an agent's 
non-knowledge of (some) future subjective contingencies? Dekel 
{\em et al\/} do not see a need for full commitment to this issue, 
but leave this point by resorting to the idea of an ``{\em as 
if\/}'' representation of their model. However, to put their model 
to the test, they call for the identification of a concrete 
Ellsberg-type example of an agent's choice behaviour which is in 
contradiction with (some of) their axioms; cf.~Dekel {\em et al\/} 
(2001) \ct[p~920]{deketal2001}.

\medskip
This challenge was met in the work by Epstein {\em et al\/} 
(2007) \ct{epsetal2007}, in which they focus on criticising Dekel 
{\em et al\/}'s additive EU representation in particular. The main 
argument Epstein {\em et al\/} give states that an economic agent 
who is aware of their incomplete knowledge of future subjective 
contingencies, and in particular is averse to this personal state 
of affairs, will feel a need to hedge against this uncertainty by 
randomisation over options available to them, thus providing a 
case of violation of Dekel {\em et al\/}'s independence axiom. In 
addition, these authors argue that the impossibility of fully 
describing all future contingencies relevant to an agent may lead 
to the failure of quantifying an agent's uncertainty about 
utilities 
``then'' in terms of just a single probability measure, as Dekel 
{\em et al\/} do in their additive EU representation. In Epstein 
{\em et al\/}'s (2007) \ct[p~359]{epsetal2007} view, Dekel 
{\em et al\/}'s model therefore precludes a consistent 
representation of incompletely known future subjective 
contingencies and ambiguity about an agent's preferences ``then''.

To overcome the conceptual problems of Dekel {\em et al\/}'s 
model --- in particular, to capture the ambiguity due to an 
agent's incomplete knowledge of future subjective contingencies, 
and their induced tendency for hedging against it ---, Epstein 
{\em et al\/} (2007) \ct{epsetal2007} propose two alternative 
axiomatic models of an agent's ex ante choice behaviour. These can 
be considered modifications of Dekel {\em et al\/}'s approach in 
the following sense. The first model maintains the assumption of
the IR axiom to hold, while the independence axiom is 
being relaxed; in the second model both the IR and the 
independence axioms are dropped, and the primitives of the model 
are extended to include random menus. The two models exhibit a 
qualitative difference as regards the status of ex post ambiguity 
that an agent finds themself exposed to ``then''. In the first 
model, an agent ``now'' expects to gain complete knowledge 
``then'' of a state realised in the meantime, i.e., before they 
choose a lottery $\beta$ from the ex-ante-preferred menu $x$; 
hence, ex post ambiguity is resolved. However, ``now'' the agent 
is uncertain about their actual preferences ``then''. In the 
second model, on the other hand, an agent ``now'' reckons that 
even ``then'' their knowledge of all relevant contingencies will 
remain incomplete, leaving their preferences ``then'' somewhat 
vague due to the lack of a complete view of all of the options 
available to them. In the present work this circumstance is 
modelled in terms of a restricted set of utility functions (over 
lotteries) with unknown likelihoods. Ex post ambiguity persists, 
making hedging against uncertainty ``then'' (and related 
potentially unfavourable outcomes) a valuable tool.

In model I, Epstein {\em et al\/} (2007) \ct{epsetal2007} 
implement an agent's need for 
hedging by following the ideas of Gilboa and Schmeidler (1989) 
\ct{gilsch1989} on uncertainty aversion in that, via introducing a 
mixing operation defined over menus, an axiomatisation 
of a multiple-priors utility representation of an agent's ex ante 
preferences is proposed. Starting from a minimal set of eight 
axioms requiring (weak) order, monotonicity, IR, non-degeneracy,\footnote{In the literature the 
names non-degeneracy and non-triviality are used synonymously for 
one of the axioms.} preference convexity, worst, 
certainty-independence, and mild continuity, the corresponding 
representation theorem states that an agent's ex ante choice 
behaviour amounts to maximising a real-valued preference 
functional $V_{MP}: \boldsymbol{X} \rightarrow \mathbb{R}$, given 
by
\be
V_{MP}(x) := \min_{\pi \in \Pi}
\int_{\boldsymbol{N}}\max_{\beta \in x}
U(\beta)\,{\rm d}\pi(U) \ ,
\ee
with $\Pi$ a (non-unique) convex and compact set of Borel 
probability measures on the space $\boldsymbol{N}$ of 
specifically normalised ex post utility functions; cf. Epstein 
{\em et al\/} (2007) \ct[p~365]{epsetal2007}.

For their model II, in order to provide a formal basis for dealing 
with persistent coarseness ``then'' of an agent's perception of 
future subjective contingencies, Epstein {\em et al\/} (2007) 
\ct{epsetal2007} enlarge the set of an agent's objects of choice 
to also include random menus of lotteries. This thus yields a set 
of Borel probability measures $\Delta(\boldsymbol{X})$ defined 
over menus in $\boldsymbol{X}$. A generic element in 
$\Delta(\boldsymbol{X})$ is denoted by $P$. Proposing a minimal 
set of six axioms comprising (weak) order, continuity, 
non-degeneracy, first-stage independence, dominance, 
and certainty reversal of order to hold, a representation theorem 
is proved for an agent's binary ex ante preference order $\succeq$ 
over random menus in $\Delta(\boldsymbol{X})$ to the extent that, 
in this model, an agent's choice behaviour corresponds to 
maximising a real-valued preference functional ${\cal V}_{PC}: 
\Delta(\boldsymbol{X}) \rightarrow \mathbb{R}$, given by
\be
{\cal V}_{PC}(P) := \int_{\boldsymbol{X}}\left[\,
\int_{{\cal K}^{cc}(\boldsymbol{N}^{*})}\max_{\beta \in x}\min_{U 
\in {\cal U}}U(\beta)\,\mu({\rm d}U)\,\right]\,{\rm d}P(x) \ ,
\ee
with $\mu \in \Delta({\cal K}^{cc}(\boldsymbol{N}^{*}))$ a Borel 
probability measure over the set ${\cal K}^{cc}(N^{*})$ of closed, 
convex and comprehensive Hausdorff-topology subsets of the compact 
space of specifically normalised ex post utility functions 
$\boldsymbol{N}^{*}$ (cf. Epstein {\em et al\/} (2007) 
\ct[p~366]{epsetal2007}), which is unique up to linear 
transformations. ${\cal U} \subset \boldsymbol{N}^{*}$ 
denotes the subset of normalised ex post utility functions 
conceived of by an agent ``now'', which, however, to them in that 
instance have unknown likelihoods as regards realisation ``then''. 
In this respect, $\boldsymbol{N}^{*} \backslash {\cal U}$ may 
be interpreted as relating to an agent's unawareness (or 
non-knowledge) ``now'' of possible subjective contingencies 
``then''.

\section{Discussion and outlook}
\lb{sec5}
Now, having discussed the state of the art in the literature, we 
should question if the formal representation of non-knowledge in 
economic theory has been satisfactory so far. Moreover, in what 
follows we sketch some promising directions of research which we 
did not address in detail in this paper.

As highlighted in section \ref{sec2}, true uncertainty and, 
thus, genuine non-knowledge about the future, are features of 
the situation in which an agent is unaware of all future 
contingencies, not (just) due to their limited ability to 
calculate, or to search for information, but due to the very 
nature of any economic system. The major insight of Knight, 
Keynes, Shackle, and 
some Post-Keynesians, was that economic systems are open and 
organic unities that are genuinely indeterminate; every decision 
situation is incomplete because it undergoes a constant change 
{\em while\/} people decide and act and, by doing so, influence 
the set of relevant variables; hence, the major characteristics of 
the decision situation --- first of all, the future states that 
are possible and conceivable --- cannot be sufficiently 
determined; they are unknown. 

We already mentioned the following concrete reasons for the 
indeterminacy of decision situations: (i)~the {\em big world 
issue\/} (i.e., the indefinite, non-exhaustive, number of possible 
future states), (ii)~the {\em endogeneity\/} of the decision 
situations, i.e., the dependence of future outcomes on 
decisions which are prepared and made in the present, and 
(iii)~the {\em social contingency\/} which is typical for economic 
systems, where the indeterminacy increases due to the dependence 
of an agent's decisions on what other agents decide. Are those 
issues adequately reflected in the {\em ambiguity\/} and {\em 
unawareness\/} approaches, which we discussed in this paper?

\subsection{Big world issue}
Savage's (1954) \ct{sav1954} axiomatisation was often criticised 
for its restrictive assumption of the ``small'' world: the list of 
possible events is presupposed to be exhaustive (though Savage 
\ct[p~16]{sav1954} himself referred to such an assumption as 
``preposterous''). Some of the follow-up concepts discussed in our 
paper differ in their treatment of this issue.

The uncertainty as ambiguity approaches we mentioned continue to 
employ Savage-type state spaces as primitives, which are 
continuous, compact, and can be partitioned into a finite number 
of mutually exclusive events, while there is an uncountable number 
of different states. Although in principle no additional structure 
is needed, some authors like Epstein and Wang (1994) 
\ct[p~206]{epswan1994} assume on the state space the existence of 
a metric and a particular (``weak convergence'') topology, 
suggesting that one can construct an indefinite 
number of different subsets of the state space; the boundaries of 
such subsets are not entirely clear. The question arises if, in 
the end, such mathematical structures make everything possible and 
thinkable, thus offering a loophole for the assumption that 
the list of possible events is not exhaustive. It is worthwhile 
mentioning here that the formal handling, but even 
more so providing compelling interpretations, of a potential 
infinitude of possibilities or states regularly proves a delicate 
issue in most (if not all) areas of applied mathematics and 
statistics; see, e.g., Hawking and Ellis (1973) \ct{hawell1973}.

In the uncertainty as unawareness models, in contrast, the big 
world issue, which relates to the state space representing the 
exogenous world, is of less importance, since here the focus of 
the analyses is on an agent's subjective state space. This 
concept, however, does not belong to the set of the primitives of 
the theory. This issue is closely related to the exogeneity vs 
endogeneity topic which we turn to discuss next.

\subsection{Endogeneity of state space}
In our view, the question of Machina (2003) \ct[p~18]{mac2003}: 
``Do individuals making choices under uncertainty face 
states of nature, or do they create them?'' remains one of the 
most crucial and controversial in decision theory. In 
Savage's concept, the state space represents nature's 
exogenous states, i.e., their emergence cannot be influenced by 
agents' decisions and actions; an agent just observes the states 
and is not an active part of the decision situation. 

In economics, as well as in the social sciences, however, there is 
increasing attention being payed to the issue of creation of 
economic reality by the actions and decisions of economic agents; 
e.g., one considers the notions of exogenous risk, performativity 
and reflexivity; cf. Danielsson and Shin (2003) \ct{danshi2003},
Danielson {\em et al\/} (2009) \ct{danetal2009}, Callon (1998) 
\ct{cal1998}, MacKenzie (2006) \ct{mac2006}, and Soros (1998) 
\ct{sor1998}. There is also an interesting movement in the 
direction of constructive decision theory, where decision-relevant 
states are not given but ``constructed by the DM 
(decision maker) in the course of deliberating about questions 
such as `How is choice A different from choice B?' and `In what 
circumstances will choice A turn out better than choice B?'''; see 
Blume {\em et al\/} (2009) \ct[p~1f]{bluetal2009}. 

Concerning the papers discussed in the article at hand, the 
works on uncertainty as ambiguity focus on the non-knowledge 
of probabilities; here, the state space remains exogenous. 
However, the unawareness literature makes an interesting and 
important move towards the conceptualisation of an endogenous 
state space: future outcomes (``then'') are contingent on
decisions made at present (``now''); the 
subjective state space is not directly observable but can be 
derived from the only variable observable: an agent's behaviour, 
or an agent's preferences (e.g., for flexibility).

However, an important question remains if, in the 
unawareness literature, we really deal  with a truly endogenous 
state space, as understood in the concepts of endogenous risk, 
performativity and reflexivity. There is already raised some 
criticism in the literature, e.g. by Sagi (2006) \ct{sag2006}, who 
is concerned about the static nature of the theoretical 
construction in 
the unawareness approaches: ``The decision maker chooses among 
menus, uncertainty over her subjective states is assumed to 
resolve and then the decision 
maker selects from the menu. However, there is no explicit 
modelling of ex-post choice and no role for consistency between 
realized tastes and tastes inferred from ex-ante preferences'', 
see Sagi (2006) \ct[p~307]{sag2006}. We also think that the 
representation of true endogeneity --- as a central determinant of 
non-knowledge --- should be dynamic: uncertainty cannot be 
resolved, as a situation constantly changes, so that the ex-post 
choice should not be modelled as a mechanic, or empty 
(i.e., predetermined) decision. The other important dynamic aspect 
is the modelling of the {\em evolution\/} of the state space 
itself (how it expands and changes); the genesis of a decision 
situation should be taken into consideration. There are some 
interesting ideas aiming at these issues, e.g. by Hayashi (2012) 
\ct{hay2012}, and Grant and Quiggin (2007) \ct{graqui2007}. The 
latter authors model the notion of discovery of the principally 
unknown space states by decision-makers. We consider this issue to 
be crucial for making further progress in the unawareness and 
non-knowledge literature.

\subsection{Social contingency}
The idea of an endogenous space state (as well as the notions of 
endogenous risk, performativity and reflexivity) goes beyond the 
subjective level of decision-making: future states are unknown 
because they are contingent on thinking, deciding and acting of 
all interconnected economic agents. This is an issue which was 
neglected in the unawareness papers presented here. However, there 
are interesting attempts to account for the social construction of 
the (subjectively perceived) state space. Here we refer to the 
work on interactive unawareness by Heifetz, Meier and Schipper 
(2006, 2008) \ct{heietal2006, heietal2008}, and on epistemic game 
theory; cf.~Brandenburger (2008) \ct{bra2008}.

\medskip
Finally, it may be noted that only a small number of authors 
proposing theoretical models of agents' choice behaviour under 
conditions of uncertainty are committed to making testable 
predictions that may be refuted in principle. This state of 
affairs conflicts with
logical positivism's view that the falsification of hypotheses by 
means of observation and/or experiment is the primary method for 
attaining meaningful progress in any empirical scientific 
disciplines; see, e.g., Popper (2002) \ct{pop2002}. Ideally, 
future research in economics and decision theory will address this 
problem more carefully.

\addcontentsline{toc}{section}{References}


\end{document}